# Automatically Tracing Imprecision Causes in JavaScript Static Analysis

Hongki Lee[a], Changhee Park[a], and Sukyoung Ryu[a]

a KAIST, South Korea

**Abstract** Researchers have developed various techniques for static analysis of JavaScript to improve analysis precision. To develop such techniques, they first identify causes of the precision losses for unproven properties. While most of the existing work has diagnosed main causes of imprecision in static analysis by manual investigation, manually tracing the imprecision causes is challenging because it requires detailed knowledge of analyzer internals. Recently, several studies proposed to localize the analysis imprecision causes automatically, but these localization techniques work for only specific analysis techniques.

In this paper, we present an automatic technique that can trace analysis imprecision causes of JavaScript applications starting from user-selected variables. Given a set of program variables, our technique stops an analysis when any of the variables gets imprecise analysis values. It then traces the imprecise analysis values using intermediate analysis results back to program points where the imprecision first started. Our technique shows the trace information with a new representation called *tracing graphs*, whose nodes and edges together represent traces from imprecise points to precise points. In order to detect major causes of analysis imprecision automatically, we present four node/edge patterns in tracing graphs for common imprecision causes. We formalized the technique of generating tracing graphs and identifying patterns, and implemented them on SAFE, a state-of-the-art JavaScript static analyzer with various analysis configurations, such as context-sensitivity, loop-sensitivity, and heap cloning. Our evaluation demonstrates that the technique can easily find 96 % of the major causes of the imprecision problems in 17 applications by only automatic detection in tracing graphs using the patterns, and selectively adopting various advanced techniques can eliminate the found causes of imprecision.

## 1 Introduction

JavaScript is the most popular language among developers according to Stack Overflow,[1] but static analysis of JavaScript applications is challenging due to its inherent dynamic and functional nature. For instance, when a dynamic property access and a first-class function call in JavaScript are combined as in `obj[e]()`, an imprecise analysis result of `e` may cause an analyzer to consider all possible method calls in the `obj` object. Then, such imprecision may lead to significant performance degradation in subsequent analysis by considering spurious calls.

Researchers have developed various static analysis frameworks [11, 12, 17] for JavaScript, and suggested techniques to improve analysis precision such as loop specialization [1, 24, 31], adaptive context-sensitivity [32], and advanced analysis domains [20, 23]. Analysis precision may be improved by repeating the following process:

1. Check if an analysis is precise enough to prove properties of interest.
2. If not,
   a. identify causes of the precision loss for the unproven properties; and
   b. use different analysis techniques or design new techniques to address the precision loss.

Existing approaches [1, 20, 23, 24, 31, 32] mainly have focused on adaptively combining existing techniques or designing new techniques in the 2-b phase relying on manual investigation for the 2-a phase. For manual investigation, however, static analyzer developers must keep track of analysis flows to identify causes of precision loss, which is expensive and requires expertise in the complex internal structure of an analyzer.

In this paper, we present a novel approach to automatically trace major causes of precision loss for given variables at certain program points when statically analyzing JavaScript programs. Throughout this paper, we regard that an analysis value is precise if it represents an abstract value that approximates a single concrete value and imprecise, otherwise; we also use the terms, *precise* (or *imprecise*) *variables* and *precise* (or *imprecise*) *program points* to denote variables and program points with precise (or imprecise) analysis values, respectively. With a given set of variables in a program, a static analyzer in our approach starts analyzing the program and stops when any of the variables gets imprecise analysis values. Then, from the points, our technique *backtracks* the imprecision sources until it finds program points where the imprecision was first introduced. It shows the trace information with a new representation, *tracing graphs*, whose nodes and edges together represent traces that show the flow of the imprecision. We also present four simple node/edge patterns to automatically detect common causes of analysis precision loss in tracing graphs.

The contributions of this paper are as follows:

- We present a new *backtracking* approach to trace causes of precision loss for given variables in JavaScript static analysis.

[1] https://insights.stackoverflow.com/survey/2018

**Listing 1** An excerpt of jQuery version 1.11

```
1 each: function(obj, callback, ...) { ...
2     var i = 0, length = obj.length ...
3     for(; i < length ; i++) { ...
4         callback.call(obj[i], i, obj[i]); ...
5     } ... }; ...
6 jQuery.each(
7     ["height", "width"],
8     function(i, name) { jQuery.cssHooks[name] = {...}; });
9 jQuery.each(
10     ["toggle", "show", "hide"],
11     function(i, name) { jQuery.fn[name] = function ... });
```

- We present *tracing graphs*, a representation that shows trace information from imprecise points to the starting points of the imprecision with four simple node/edge patterns for common causes of precision loss.
- Our experimental evaluation with 12 versions of jQuery,[2] the most popular JavaScript library, and 5 JavaScript benchmarks from a previous work [13], demonstrates that the technique can easily find 144 out of 150 major causes (96 %) in 17 applications by only automatic detection. It also shows that a new analysis using our technique outperforms a state-of-the-art JavaScript analyzer [24] in analyzing 13 out of 17 applications with higher analysis precision.

## 2 Overview

In this section, we show difficulties of tracing causes of precision loss in static analysis of JavaScript programs with a concrete example, and present our technique to detect the problem automatically.

### 2.1 Motivation

Listing 1 shows a motivating example, which is an excerpt of jQuery version 1.11; we omit unimportant parts in describing our technique using ellipses. This excerpt defines the jQuery.each method (lines 1–5) and calls it twice (lines 6 and 9) with two arguments, an array of string elements and a callback function. The method iterates over each element of the array in a loop (line 3) and calls the callback function with the element and its index in the array (line 4); callback.call(obj[i], i, obj[i]) calls callback with the receiver object, obj[i], and two arguments, i and obj[i]. Then, the two callback functions (lines 8 and 11) add two properties with the names "height" and "width" to the object jQuery.cssHooks (line 8), and three properties with the names "toggle", "show", and "hide" to the object jQuery.fn (line 11), respectively.

[2] https://jquery.com

Now, let us assume that we use a static analyzer with flow-sensitivity, which takes the order of program statements into account, context-insensitivity, which does not distinguish different calling contexts of the same function, and loop-insensitivity, which does not distinguish iterations of loops. Then, two array objects at lines 7 and 10 at the two calls of the jQuery.each method are merged into the abstract value of obj at the entry of the method. Note that *merging* abstract values means generating a new abstract value that over-approximates them. Then, the possible abstract value of i in the loop body (line 4) includes concrete numbers between 0 and 2. With these two imprecise values of obj and i combined, obj[i] may point to any string values in the two arrays ("height", "width", "toggle", "show", and "hide"), and this imprecise value is passed to two callback functions (lines 8 and 11) through the name parameters. Consequently, jQuery.cssHooks and jQuery.fn objects are updated (lines 8 and 11) with five possible string names of properties from the abstract value of name. To improve the precision of the jQuery.cssHooks and jQuery.fn objects, we would like to find and resolve causes of the precision loss for the name variables.

However, manually tracing the causes of the precision loss is not only time-consuming and tedious but also challenging because it requires a fair knowledge of the analyzer internals. For instance, consider tracing the imprecise variable name at line 11 that points to an abstract value indicating five concrete string values. First, we have to find out where the imprecise value comes from inside the function. While it is easy to see that name is a parameter of the function at line 11 in this simple case, it is generally difficult to find the imprecision source. We often have to find out all program points that may affect the precision of the abstract value tracing back the program flows, which may involve tracing several other variables and repeating this process several times more. Next, by inspecting the call graph of the code in listing 1, we can discover that the imprecise value of the name parameter comes from the abstract value of obj[i] at the function call (line 4). Now, we should trace the imprecise value of obj[i] and we may consider the following possibilities for the imprecision:

1. the obj or the i variable may have an imprecise value; and
2. the obj and i variables have precise values but obj[i] has an imprecise value.

Consider that obj has an imprecise value that may point to one of two arrays (lines 7 and 10) and i also has an imprecise value that may represent one number between 0 and 2. Then, we should trace the imprecision sources of obj and i using the same process; we can discover that the abstract value of obj comes from the first parameters of the two calls, which have precise values that are arrays of string constants, and that the abstract value of i is a precise value 0 before the loop. Note that we trace the variables obj and i first because obj[i] cannot be precise if those variables have imprecise values. At this point, we can identify where the imprecision first occurred: the precise array values at the calls (lines 6 and 9) were merged as an imprecise value of obj at the entry of the each method and the abstract values of i from different iterations of the loop were merged at the head of the loop. In this case, we can resolve the imprecision by applying context-sensitivity and loop-specialization techniques that can distinguish two calls and the loop iterations, respectively. Note that if precise values for obj and i do not resolve the imprecision of name, we have to consider the second possibility (tracing the imprecise value of obj[i]) to identify the imprecision source.

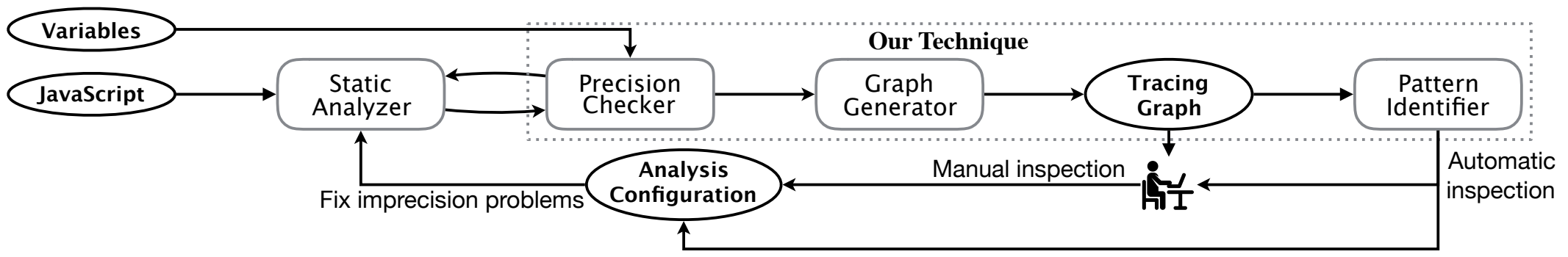

**Figure 1** Overview of our technique and its applications

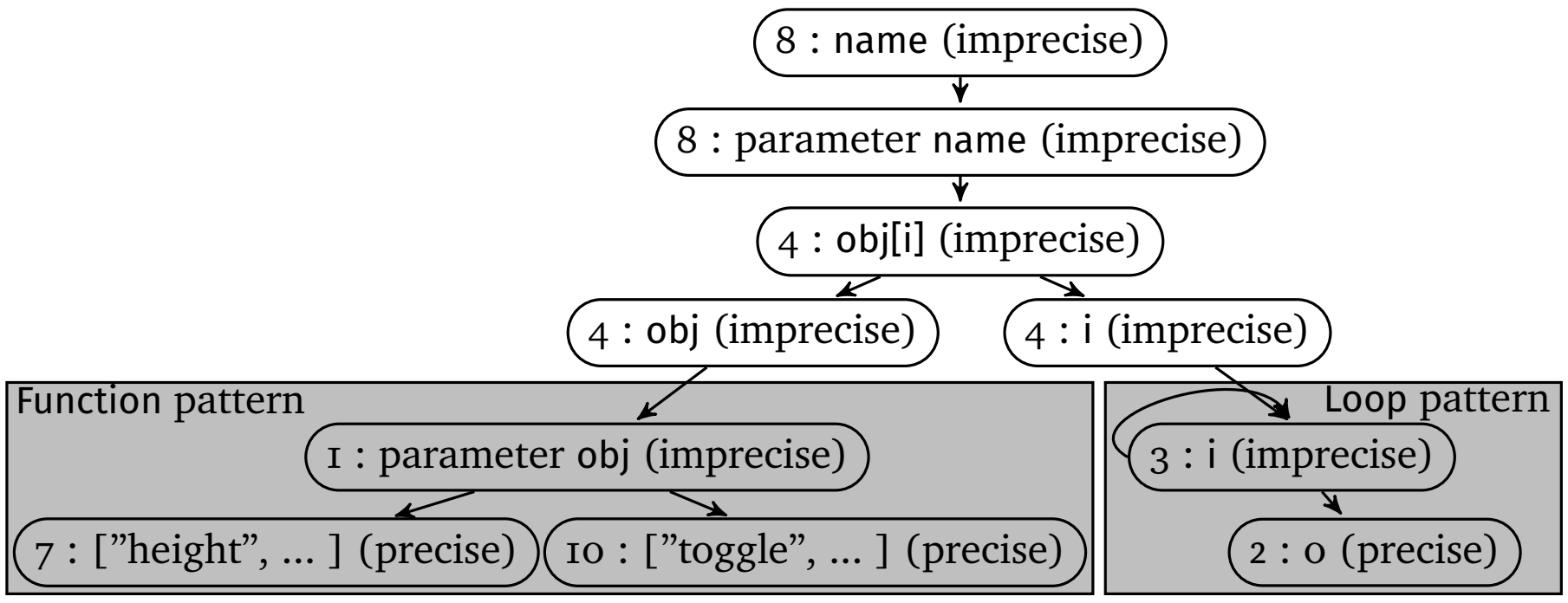

**Figure 2** A tracing graph for name at line 8 in listing 1

## 2.2 Our Technique

We propose a technique that makes the tracing process fully automatic, and thus enables users to easily find main causes of imprecision. Figure 1 illustrates how the technique combined with a static analyzer can improve the precision of JavaScript analysis. The technique consists of three components: Precision Checker, Graph Generator, and Pattern Identifier. When a user provides a JavaScript program and a set of variables whose abstract values need to be checked for precision to Static Analyzer and Precision Checker, Precision Checker periodically checks the precision of the abstract values of the given variables while the static analyzer analyzes the program, and it stops the analysis when any of the variables gets imprecise values. Then, it passes intermediate analysis results and the imprecise program points to Graph Generator. From the imprecise program points, Graph Generator traces back the program flows and produces a graph called a *tracing graph* that shows the trace information of the imprecise values in a concise way. Using the generated tracing graph, the user can manually inspect the flows of the imprecise values to identify main causes of the imprecision.

For instance, figure 2 shows a simplified tracing graph for the imprecision of the name variable at line 8 in listing 1. Each node contains three pieces of information: 1) a source location, which is a line number in listing 1, 2) an expression that the technique keeps track of, which is a variable, a property load, or a constant, and 3) whether its abstract value at the location is precise or not. For example, the top-most node denotes that the name variable has an imprecise value at line 8. The edges denote flows of the imprecise values through important program points where analysis precision losses may happen. Note that the node for obj[i] at line 4 has two successors, one for obj and another for i. Recall the two possibilities that we described in the previous subsection to trace the imprecision of obj[i]. When all the two possibilities, variables obj and i and a property load obj[i], are available, we decide to trace one rather than both. If a variable obj or i has an imprecise value, a property load obj[i] becomes imprecise, so that the technique gives the highest priority to the first case by heuristics and traces the imprecision sources for obj and i; it considers the second case only when both obj and i have precise values. Thus, in this example, the tracing graph has two successors of the node for obj[i] at line 4 by first case because both obj and i have imprecise values.

In order to discover main causes of imprecision in tracing graphs, users can utilize Pattern Identifier that can automatically detect imprecision causes. For example, for the graph in figure 2, Pattern Identifier can detect the Function pattern (left shaded box), where all the leaf nodes with precise values that share a common predecessor point to the callsites of the jQuery.each method and the predecessor points to the entry of the method at line 1. For this case, it reports that one of the main causes of the analysis imprecision is due to the precision loss at the function entry at line 1. For the same graph, Pattern Identifier can detect another pattern called the Loop pattern (right shaded box), where one node points to a loop head with a cyclic edge and its successor is a leaf node with a precise value. Then, it reports that one of the main causes of the imprecision is due to the precision loss in the loop at line 3. We present two more patterns that Pattern Identifier can detect and detection rules for them in section 5.

Finally, by manually or automatically inspecting results, the user can improve the precision of subsequent analysis by updating Analysis Configuration with advanced analysis techniques that address the detected imprecision causes. With a new configuration, the static analysis can analyze the program to get more precise results.

## 3 Base Analyzer

We present an overview of a base analyzer that we extend to present our technique in the next section. We assume a static analyzer based on the abstract interpretation framework [6] using a standard iterative worklist algorithm [14], which supports flow-sensitivity and $k$-CFA [26, 27] (distinguishing calling contexts of functions using call history information).

We also assume that the analyzer represents JavaScript programs as Control Flow Graphs (CFGs). Figure 3 summarizes the abstract syntax of CFGs for a JavaScript core language, which is sufficient to explain the main idea of our technique. A CFG consists of a set of nodes $\wp(\mathcal{N}_c)$ and a set of edges $\wp(\mathcal{E}_c)$. A node $n = \langle f, i, \hat{c} \rangle \in \mathcal{N}_c$ is a triple of a function id, which is a unique number for each function, a CFG instruction, and a context during analysis. An edge $\langle n_1, n_2 \rangle \in \mathcal{E}_c$ is a pair of two nodes representing a potential program flow from $n_1$ to $n_2$. An expression $e$ is one of a constant value $c$, a variable $x$, an object property load $x_1[x_2]$, and a binary operation $x_1 \oplus x_2$. We leave analysis context $\hat{\mathcal{C}}$ abstract. The concrete semantics of each instruction is as follows:

- entry$[x_1,\dots,x_n]$ denotes the entry point of a function with parameters $x_1,\dots,x_n$ whose values are passed from the array of arguments $x_{\text{args}}$ in the corresponding call instruction call$[x_{\text{f}}, x_{\text{this}}, x_{\text{args}}]$.
- exit denotes the end of a function; the return value of the function is propagated to $x$ of the after-call$[x]$ instruction in its caller.
- alloc$[x]$ allocates an address for an empty object in a heap and assigns the address to $x$.
- function$[x,f]$ creates a function value with the function id $f$; because, in JavaScript, a function is also an object, this instruction allocates an address for the function object and assigns the address to $x$. In addition to handling non-local variables in $f$, the function object includes the lexical scope information of the current scope.
- write-var$[x,e]$ evaluates the expression $e$ and stores the result into $x$.
- write-prop$[x_1[x_2],x_3]$ updates a property of the object that $x_1$ points to with the property name of the value of $x_2$ by writing the value of $x_3$ to the property.
- call$[x_{\text{f}}, x_{\text{this}}, x_{\text{args}}]$ connects an inter-procedural edge from this node to the entry node of the function that $x_{\text{f}}$ points to, and updates this and arguments (an array that stores arguments) of a callee function with the values $x_{\text{this}}$ and $x_{\text{args}}$.
- after-call$[x]$ denotes a return point from a callee function of the corresponding call with the return value stored in $x$ that the callee passed.
- return$[x]$ stores the value of $x$ as the result of the current function.
- cond$[x_1 \oplus x_2]$ denotes a head node of conditional branches with a conditional expression, for which we consider only a binary operation, $x_1 \oplus x_2$, for brevity.
- builtin$[c, x_{\text{this}}, x_{\text{args}}]$ denotes a JavaScript built-in function with the kind information $c$. Depending on $c$, it performs different semantics and its return value can be differentiated by $x_{\text{this}}$ and $x_{\text{args}}$.
- skip just propagates states to next nodes without any changes.

| | | | | | | |
|---|---|---|---|---|---|---|
| (CFG) | $\mathcal{G}_c =$ | $\wp(\mathcal{N}_c) \times \wp(\mathcal{E}_c)$ | (Function Id) | $f$ | $\in$ | $\mathcal{F} = \{0,1,2,...\}$ |
| (CFG Node) | $n \in \mathcal{N}_c =$ | $\mathcal{F} \times \mathcal{I} \times \hat{\mathcal{C}}$ | (Variable) | $x$ | $\in$ | $Var$ |
| (CFG Edge) | $\langle n_1, n_2 \rangle \in \mathcal{E}_c =$ | $\mathcal{N}_c \times \mathcal{N}_c$ | (Expression) | $e$ | $::=$ | $c \mid x \mid x[x] \mid x \oplus x$ |
| (Analysis Context) | $\hat{c} \in \hat{\mathcal{C}}$ | | | | | |
| (Instruction) | $i \in \mathcal{I}$ ::= | entry$[x,\dots,x]$ | \| exit | | | \| alloc$[x]$ |
| | \| | function$[x,f]$ | \| write-var$[x,e]$ | | | \| write-prop$[x[x],x]$ |
| | \| | call$[x,x,x]$ | \| after-call$[x]$ | | | \| return$[x]$ |
| | \| | cond$[x \oplus x]$ | \| builtin$[c]$ | | | \| skip |

**Figure 3** Control flow graph of a core language

One JavaScript statement may be transformed into multiple CFG instructions. Figure 4 illustrates how a JavaScript method call statement "$y$ = $e$.foo($x$)" is transformed into multiple instructions where $x$ and $y$ are variables, $e$ is an expression, and foo is a string constant. Note that the evaluation results of all the expressions in the call statement are stored into some variables first, and the actual function call is performed only with the values of the variables as in call[$x_f$,$x_{\text{this}}$,$x_{\text{args}}$]. The nodes for the call and after-call instructions are connected by inter-procedural edges (not shown in figure 4) to the CFG of the $e$.foo method and the return value of the method is stored in $y$ that the after-call instruction has.

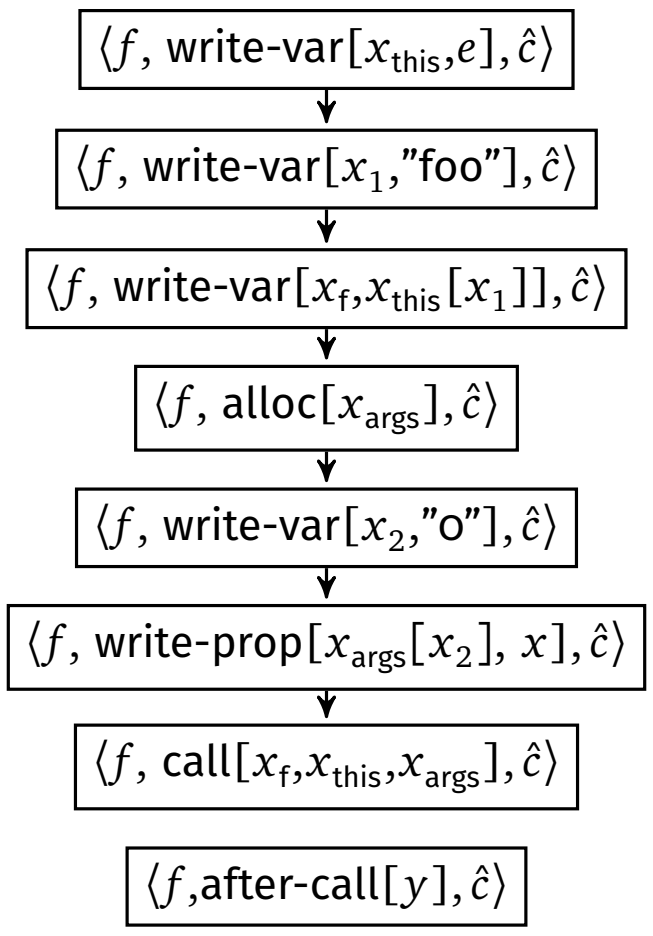

**Figure 4** Transformed CFG from a JavaScript statement

On CFGs, the base analyzer computes analysis results using the abstract semantics of each instruction, which approximates the result of its concrete semantics in a sound way. For a given node $\langle f, i, \hat{c}\rangle$ and a variable $x$ or an object property load $x_1[x_2]$ in a given program, a summary function $\hat{\mathcal{T}}$ gives the analysis value of the variable or the object property before evaluation of the instruction $i$ in the function $f$ with the context $\hat{c}$. After computing a fixpoint of the abstract semantics of a given program, the summary function $\hat{\mathcal{T}}$ represents the analysis result of the program. However, when the analysis result of a variable or a property load is imprecise, it may lead to an imprecision problem discussed in section 2. In the next section, we present how our technique traces the source of such an imprecision back to its first occurrence point.

## 4 Imprecision Tracing in Static Analysis

Among three components of our technique, we explain Precision Checker and Graph Generator in this section and Pattern Identifier in the next section.

### 4.1 Precision Checker

When a user provides a program and a set of variables, Precision Checker periodically checks the precision of the abstract values for the given variables while the base analyzer analyzes the program. It terminates the analysis when any of the variables gets imprecise values. We implemented it by modifying the worklist algorithm [14] of the analyzer to compute a fixpoint for a summary function $\hat{\mathcal{T}}$.

We assume that a user provides $I$, a set of variables and their program points as CFG nodes: $I \in \wp(\mathcal{N}_c \times Var)$. Then, at each iteration of the fixpoint computation, the

$$
\begin{array}{llll}
\text{(Tracing Graph)} & \mathcal{G}_t = \wp(\mathcal{N}_t) \times \wp(\mathcal{E}_t) & \text{(Object Location)} & \hat{l} \in \hat{Loc} \\
\text{(Tracing Edge)} & \mathcal{E}_t = \mathcal{N}_t \times \mathcal{N}_t & \text{(Property Name)} & s \in String \\
\text{(Tracing Node)} & t \in \mathcal{N}_t ::= \mathsf{Pre}_{\mathsf{Var}}(n,x) \mid \mathsf{Impre}_{\mathsf{Var}}(n,x) \mid \mathsf{Pre}_{\mathsf{Prop}}(n,\hat{l},s) \mid \mathsf{Impre}_{\mathsf{Prop}}(n,\hat{l},s) & &
\end{array}
$$

**Figure 5** Definition of tracing graphs

**Algorithm 1:** Main algorithm of Graph Generator

**Input:** *impreVars*, $\hat{\mathcal{T}}$
**Output:** $G$
$worklist := impreVars,\ G := \langle\emptyset,\emptyset\rangle$ ;
**while** !*worklist.isEmpty*() **do**
  $t := worklist.pop()$ ;
  **if** $t$ *contains a node with either* entry *or* after-call **then**
    $ts :=$ TRACEINTER$(t,\hat{\mathcal{T}})$ ;
  **else**
    $ts :=$ TRACEINTRA$(t,\hat{\mathcal{T}})$ ;
  $worklist := (worklist \cup \text{filterPreciseNode}(ts)) \setminus G.nodes$ ;
  $G.nodes := G.nodes \cup ts$ ;
  $G.edges := G.edges \cup \{\langle t,t'\rangle \mid t' \in ts\}$ ;
**return** $G$ ;

precision checker tests whether any variables from $I$ get imprecise values and collects them as starting points for tracing imprecision as follows:

$$impreVars := \{\mathsf{Impre}_{\mathsf{Var}}(n,x) \mid \langle n,x\rangle \in I \wedge impre?(n,x,\hat{\mathcal{T}})\}$$

where $impre?(n,x,\hat{\mathcal{T}})$ checks whether the abstract value of $x$ at node $n$ using the intermediate result $\hat{\mathcal{T}}$ produces an imprecise value; it returns true if the abstract value of $x$ approximates two or more concrete values and false, otherwise. The set *impreVars* collects tracing graph nodes $\mathsf{Impre}_{\mathsf{Var}}(n,x)$ explained in the next subsection, which denotes that the abstract value of $x$ is imprecise at node $n$. When the worklist algorithm finds that *impreVars* is not empty, it immediately stops the analysis and passes the intermediate result $\hat{\mathcal{T}}$ and the set of starting tracing graph nodes *impreVars* to Graph Generator to produce a tracing graph.

### 4.2 Tracing Graph Generation

We represent tracing information to detect the sources of imprecise results using *tracing graphs* defined in figure 5. A tracing graph consists of a set of tracing nodes $\wp(\mathcal{N}_t)$ and a set of tracing edges $\wp(\mathcal{E}_t)$. A tracing node $t \in \mathcal{N}_t$ denotes one of four cases: $\mathsf{Pre}_{\mathsf{Var}}(n,x)$ and $\mathsf{Pre}_{\mathsf{Prop}}(n,\hat{l},s)$ denote that variable $x$ and property $s$ of abstract object at $\hat{l}$ have precise values at node $n$, respectively, and $\mathsf{Impre}_{\mathsf{Var}}(n,x)$ and $\mathsf{Impre}_{\mathsf{Prop}}(n,\hat{l},s)$ denote that $x$ and property $s$ of abstract object at $\hat{l}$ have imprecise values at node $n$, respectively.

Algorithm 1 describes the main algorithm to generate a tracing graph in Graph Generator. It takes the set of starting nodes *impreVars* and the intermediate result $\hat{\mathcal{T}}$ as

$$
\textsc{TraceIntra}(t,\hat{\mathscr{T}}) = \quad \textit{where } t = \mathsf{Impre}_{\mathsf{Var}}(n,x) \wedge n_i = \textit{idom}(n)
$$

$$
\begin{cases}
\{\mathsf{Impre}_{\mathsf{Var}}(n_i,x)\} \quad \textit{if } (x \text{ is merged at } n_i) \wedge n_i = \langle _, \mathsf{skip}, _ \rangle \\
\hline
\{\mathsf{Impre}_{\mathsf{Var}}(n',w) | \langle n',w \rangle \in nvs_1 \cup nvs_2\} \\
\textit{if} \begin{cases} (x \text{ is merged at } n) \wedge nvs_1 = \{\langle n_p,x \rangle | n_p \in \mathrm{getPreds}(n) \wedge \textit{impre?}(n_p,x,\hat{\mathscr{T}})\} \\ \wedge\, n_i = \langle _, \mathsf{cond}[y \oplus z], _ \rangle \wedge nvs_2 = \{\langle n_i,w \rangle | w \in \{y,z\} \wedge \textit{impre?}(n_i,w,\hat{\mathscr{T}})\} \end{cases} \\
\hline
\{\mathsf{Impre}_{\mathsf{Var}}(n_i,w) | w \in vs\} \\
\textit{if } \{\, n_i = \langle _, \mathsf{write\text{-}var}[x,x_1[x_2]], _ \rangle \wedge vs = \{w | w \in \{x_1,x_2\} \wedge \textit{impre?}(n_i,w,\hat{\mathscr{T}})\} \wedge 0 < |vs| \\
\hline
\{\mathsf{Impre}_{\mathsf{Prop}}(n_i,\hat{l},s)\} \\
\textit{if} \begin{cases} n_i = \langle _, \mathsf{write\text{-}var}[x,x_1[x_2]], _ \rangle \wedge \neg\textit{impre?}(n_i,x_1,\hat{\mathscr{T}}) \wedge \neg\textit{impre?}(n_i,x_2,\hat{\mathscr{T}}) \\ \wedge\, \textit{impre?}(n_i,x_1[x_2],\hat{\mathscr{T}}) \wedge \{\hat{l}\} = \hat{\mathscr{T}}(n_i,x_1) \wedge \{s\} = \hat{\mathscr{T}}(n_i,x_2) \end{cases}
\end{cases}
$$

**Figure 6** Partial definition of TraceIntra

its inputs. It first initializes *worklist* and the output tracing graph $G$ as *impreVars* and the empty graph. From tracing nodes in *worklist*, it finds next imprecise tracing nodes to trace by the intra-procedural (TraceIntra) and the inter-procedural (TraceInter) tracing rules. Depending on CFG instructions in the current nodes, it decides which rule to apply between TraceIntra and TraceInter. If the current node contains a node for entry or after-call, it performs inter-procedural tracing by TraceInter; otherwise, it does intra-procedural tracing by TraceIntra. When one of the tracing functions generates a set of tracing graph nodes to trace further for representing the imprecision flow, it adds only those that denote imprecision and do not already exist in $G$ into *worklist*; filterPreciseNode($ts$) filters out precise tracing graph nodes such as $\mathsf{Pre}_{\mathsf{Var}}$ and $\mathsf{Pre}_{\mathsf{Prop}}$ from the set of newly generated nodes $ts$. Finally, it updates the nodes and edges of $G$ with the new nodes that the tracing functions generated. It repeats the process until no more imprecise tracing nodes remain in *worklist*. Figure 6 and figure 7 show partial definitions of TraceIntra and TraceInter, respectively.

### 4.2.1 Intra-procedural Tracing

TraceIntra($t,\hat{\mathscr{T}}$) backtracks a control flow through intra-procedural edges using $\hat{\mathscr{T}}$ from a tracing node $t$ that has an imprecise variable or property access $e$. It traces the control flow until it finds either 1) the definition of $e$, write-var or write-prop assigning an abstract value to $e$, or 2) a *join point* where the abstract values of $e$ from multiple control flows are merged. For the first case, if a precise value is assigned to $e$, it generates a precise tracing node, $\mathsf{Pre}_{\mathsf{Var}}$ or $\mathsf{Pre}_{\mathsf{Prop}}$, for $e$ and ends tracing $e$; otherwise, it generates an imprecise tracing node, $\mathsf{Impre}_{\mathsf{Var}}$ or $\mathsf{Impre}_{\mathsf{Prop}}$, for some variable or property access $e'$ that affects the imprecision of $e$ and keeps tracing $e'$. Consider the following simplified CFG for example:

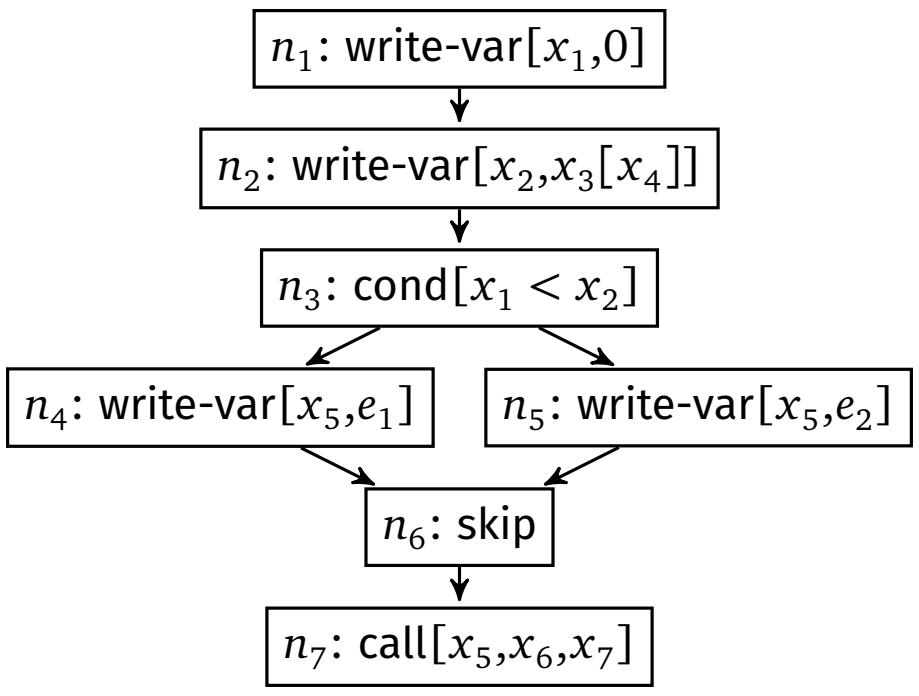

where we omit function ids and contexts but annotate node numbers like $n_1$ and $n_2$ in the CFG nodes for presentation for brevity. Assume that we trace the imprecise value of $x_5$ that may have multiple callees at $n_7$. Then, from $n_7$, TraceIntra first traces the imprecise value back to $n_6$ by the first rule in figure 6 because $n_6$ is a join point of $x_5$ with two definitions coming from $n_4$ and $n_5$. The rule computes the *immediate dominator* of $n_7$ (denoted as $idom(n_7)$), which indicates a unique closest node to $n_7$ among common nodes that every path from the entry node to $n_7$ in intra-procedural flows must go through before reaching $n_7$, to generate a tracing node $\mathsf{Impre}_{\mathsf{Var}}(n_6,x_5)$.

Next, the imprecise value of $x_5$ at $n_6$ may happen because either: 1) the predecessors $n_4$ and $n_5$ of $n_6$ propagated imprecise values, or 2) they have precise values for $x_5$ but the abstract values were merged at $n_6$ due to an unknown condition of $x_1 < x_2$ at $n_3$. The second rule handles both cases. It gets the predecessors and the branch head node for $n_6$ using getPreds($n_6$) and $idom(n_6)$, respectively, and generates imprecise tracing graph nodes $\mathsf{Impre}_{\mathsf{Var}}(n_4,x_5)$, $\mathsf{Impre}_{\mathsf{Var}}(n_5,x_5)$, $\mathsf{Impre}_{\mathsf{Var}}(n_3,x_1)$, and $\mathsf{Impre}_{\mathsf{Var}}(n_3,x_2)$ if variables in their program points have imprecise values. Assuming that $x_5$ at $n_4$ and $n_5$ has already precise values, then TraceIntra ends tracing not only the abstract values of $x_5$ at $n_4$ and $n_5$ but also the abstract value of $x_1$ at $n_3$ because $x_1$ has a precise value 0 from the assignment at $n_1$. Note that the rules to generate $\mathsf{Pre}_{\mathsf{Var}}$ and $\mathsf{Pre}_{\mathsf{Prop}}$ for variables can be obtained using $\{\mathsf{Pre}_{\mathsf{Var}}(n', x) | n' \in \text{getPreds}(n) \land \neg \textit{impre?}(n', x, \hat{\mathcal{T}})\} \cup \{\mathsf{Pre}_{\mathsf{Var}}(n_i, w) | w \in \{y, z\} \land \neg \textit{impre?}(n_i, w, \hat{\mathcal{T}})\}$ by negating *impre?* in the rules for $\mathsf{Impre}_{\mathsf{Var}}$ and $\mathsf{Impre}_{\mathsf{Prop}}$. Therefore, if $x_5$ has a precise value at $n_4$ but it has an imprecise value at $n_5$, this rule generates $\mathsf{Pre}_{\mathsf{Var}}(n_4,x_5)$ and $\mathsf{Impre}_{\mathsf{Var}}(n_5,x_5)$.

For $x_2$, if it has an imprecise value, its abstract value comes from the imprecise value of $x_3[x_4]$ at $n_2$. In this case, TraceIntra considers the following two cases in the order of heuristic priorities: 1) $x_3$ or $x_4$ has an imprecise value (the third rule) and 2) both $x_3$ and $x_4$ have precise values but $x_3[x_4]$ has an imprecise value (the fourth rule). In the fourth rule, $\textit{impre?}(n, x_3[x_4], \hat{\mathcal{T}})$ checks whether $x_3[x_4]$ in $\hat{\mathcal{T}}$ produces an imprecise value. For each case, TraceIntra generates $\mathsf{Impre}_{\mathsf{Var}}(n_2, x_3)$, $\mathsf{Impre}_{\mathsf{Var}}(n_2, x_4)$, and $\mathsf{Impre}_{\mathsf{Prop}}(n_2, \hat{l}, s)$, respectively, where $\hat{l}$ is the location of the object that $x_3$ points to and $s$ is the concrete string value pointed by $x_4$ that is used as a property name. Then, TraceIntra keeps tracing the imprecision sources of the new targets.

#### 4.2.2 Inter-procedural Tracing

TraceInter($t$,$\hat{\mathcal{T}}$) backtracks an inter-procedural flow from a node pointed by the input tracing graph node $t$ and generates new tracing graph nodes to keep tracing

$$
\textsc{TraceInter}(t, \hat{\mathscr{T}}) = \ \mathit{where}\ t = \mathsf{Impre}_{\mathsf{Var}}(n, x) \wedge \mathit{preds} = \mathrm{getPreds}(n)
$$

$$
\begin{cases}
\{\mathsf{Impre}_{\mathsf{Var}}(n', y) \mid \langle n', y\rangle \in \mathit{nvs}\} \\
\mathit{if} \begin{cases} n = \langle _, \mathsf{entry}[x_1, ..., x_n], _\rangle \wedge x = \mathsf{this} \\ \wedge\ \mathit{nvs} = \{\langle n', y\rangle \mid n' = \langle _, \mathsf{call}[_, y, _], _\rangle \wedge n' \in \mathit{preds} \wedge \mathit{impre?}(n', y, \hat{\mathscr{T}})\} \end{cases} \\
\hline
\{\mathsf{Impre}_{\mathsf{Prop}}(n', \hat{l}, *\mathsf{scope}) \mid \langle n', \hat{l}\rangle \in \mathit{nls}\} \cup \{\mathsf{Impre}_{\mathsf{Var}}(n', x) \mid n' \in \mathit{ns}\} \\
\mathit{if} \begin{cases} n = \langle _, \mathsf{entry}[x_1, \ldots, x_n], _\rangle \\ \wedge\ \mathit{nls} = \{\langle n', \hat{l}\rangle \mid \hat{l} = \mathrm{getFunc}(n) \wedge n' \in \mathit{preds} \wedge \mathit{impre?}(n', \langle \hat{l}, *\mathsf{scope}\rangle, \hat{\mathscr{T}})\} \\ \wedge\ \mathit{ns} = \{\ n' \mid \hat{l} = \mathrm{getFunc}(n) \wedge n' \in \mathit{preds} \wedge \neg \mathit{impre?}(n', \langle \hat{l}, *\mathsf{scope}\rangle, \hat{\mathscr{T}}) \wedge \mathit{impre?}(n', x, \hat{\mathscr{T}})\} \end{cases} \\
\hline
\{\mathsf{Impre}_{\mathsf{Var}}(n', *\mathsf{ret}) \mid n' \in \mathit{ns}\} \\
\mathit{if} \ \{\ n = \langle _, \mathsf{after\text{-}call}[x], _\rangle \wedge \mathit{ns} = \{n' \mid n' \in \mathit{preds} \wedge \mathit{impre?}(n', *\mathsf{ret}, \hat{\mathscr{T}})\}
\end{cases}
$$

**Figure 7** Partial definition of TRACEINTER

analysis imprecision in the inter-procedural predecessors of the current CFG node. Note that analysis imprecision can flow from a caller function to a callee function and vice versa via 1) arguments including the abstract value of the this variable at the callee, 2) shared non-local variables that are defined outside the scope of the caller and the callee, and 3) a return variable.

For call nodes, consider the following inter-procedural flows from call instructions to the entry of a callee function:

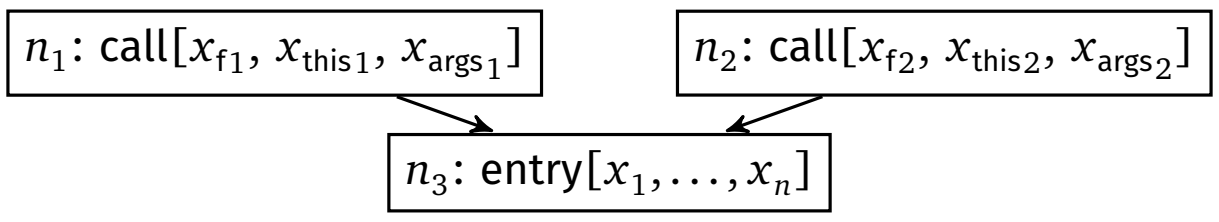

where two call instructions at $n_1$ and $n_2$ invoke the same function with the same context. Let us assume that TRACEINTRA has traced the analysis imprecision until $n_3$ and generated the tracing graph node, $\mathsf{Impre}_{\mathsf{Var}}(n_3, \mathsf{this})$, meaning that the this variable has an imprecise value at $n_3$. Then, algorithm 1 calls TRACEINTER for the inter-procedural tracing. The first rule in figure 7 generates two nodes, $\mathsf{Impre}_{\mathsf{Var}}(n_1, x_{\mathsf{this}1})$ and $\mathsf{Impre}_{\mathsf{Var}}(n_2, x_{\mathsf{this}2})$, if both $x_{\mathsf{this}1}$ and $x_{\mathsf{this}2}$ that pass abstract values for this are imprecise at $n_1$ and $n_2$, respectively. Tracing other parameters at entry nodes is performed in a similar manner.

On the other hand, consider tracing imprecise values of non-local variables. Assume that a variable $x$ at $n_3$ is a non-local variable with an imprecise value. Note that because JavaScript supports lexically-scoped higher-order functions, the same function at different contexts may be defined several times with different lexical scope information; we assume that the internal $*$scope property of a function object contains the lexical scope information that has values of non-local variables in the function. Thus, the imprecision of $x$ may be due to imprecise scope information that abstracts multiple concrete scopes. In this case, TRACEINTER considers the following two possibilities in the order of the priorities determined by heuristics: 1) the lexical scope information of the current scope is imprecise and 2) the lexical scope information is precise but $x$ itself has an imprecise value. For the first case, *nls* in the second rule in figure 7 is used to generate either $\mathsf{Impre}_{\mathsf{Prop}}(n_1, \hat{l}, *\mathsf{scope})$ or $\mathsf{Impre}_{\mathsf{Prop}}(n_2, \hat{l}, *\mathsf{scope})$, or both of them, where $\mathrm{getFunc}(n)$ gives an abstract function object enclosing $n$, and the internal $*$scope property of the current function object denoted by the location $\hat{l}$ has the lexical

$$
\textsc{Pattern}(t,E) = \begin{cases}
\textsf{Function} & \\
\quad \textit{if} \begin{cases} \big(t = \textsf{Impre}_{\textsf{Var}}(\langle _, \textsf{entry}[x_0,\ldots,x_n], _\rangle, _) \lor t = \textsf{Impre}_{\textsf{Prop}}(\langle _, \textsf{entry}[x_0,\ldots,x_n], _\rangle, _, _)\big) \\ \land \underset{\langle t,t'\rangle \in E}{\forall} \big(t' = \textsf{Pre}_{\textsf{Var}}(\langle _, \textsf{call}[_,_,_], _\rangle, _) \lor t' = \textsf{Pre}_{\textsf{Prop}}(\langle _, \textsf{call}[_,_,_], _\rangle, _, _)\big) \end{cases} \\
\hline
\textsf{Loop} & \\
\quad \textit{if} \ \{ t = \textsf{Impre}_{\textsf{Var}}(\langle _, \textsf{cond}[y \oplus z], _\rangle, x) \land x \in \{y,z\} \land t \rightsquigarrow t \land \langle t, \textsf{Pre}_{\textsf{Var}}(_, x)\rangle \in E \\
\hline
\textsf{HeapClone} & \\
\quad \textit{if} \begin{cases} \big(\ t = \textsf{Impre}_{\textsf{Prop}}(\langle _, \textsf{write-prop}[_,_], _\rangle, _, _)\rangle \lor t = \textsf{Impre}_{\textsf{Prop}}(\langle _, \textsf{alloc}[_], _\rangle, _, _)\rangle\big) \\ \land \underset{\langle t,t'\rangle \in E}{\forall} \big(\ t' = \textsf{Pre}_{\textsf{Var}}(\langle _,_,_\rangle, _) \lor t' = \textsf{Pre}_{\textsf{Prop}}(\langle _,_,_\rangle, _, _)\big) \end{cases} \\
\hline
\textsf{Model} & \\
\quad \textit{if} \begin{cases} \big(t = \textsf{Impre}_{\textsf{Var}}(\langle _, \textsf{builtin}[c], _\rangle, {*}\textsf{ret}) \land \langle t, \textsf{Pre}_{\textsf{Var}}(n', _)\rangle \in E\big) \\ \lor \big(t = \textsf{Impre}_{\textsf{Prop}}(\langle _, \textsf{entry}[x_0,\ldots,x_n], _\rangle, _, _) \land \langle t, _\rangle \notin E\big) \end{cases} \\
\hline
\textsf{Unknown} & \textit{otherwise}
\end{cases}
$$

**Figure 8** Node/edge pattern detection rules

information when the function was defined. For the second case, the rule simply generates either $\textsf{Impre}_{\textsf{Var}}(n_1,x)$ or $\textsf{Impre}_{\textsf{Var}}(n_2,x)$, or both of them to keep tracing the imprecise value of $x$ at the callers using $ns$.

For exit nodes, consider the following inter-procedural flows to the return site of the same caller with an abstract value stored in $x_{\textsf{ret}}$:

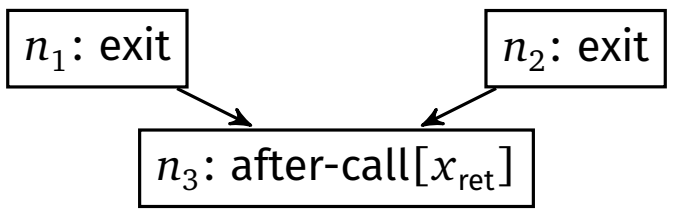

When a function passes its result to its caller, we assume that the result is stored to an internal variable ∗ret in the base analyzer. In this case, when TraceIntra traces the imprecise value of $x_{\textsf{ret}}$ at $n_3$, the last rule generates $\textsf{Impre}_{\textsf{Var}}(n_3,{*}\textsf{ret})$. Then, subsequently, TraceInter generates $\textsf{Impre}_{\textsf{Var}}(n_1,{*}\textsf{ret})$ and $\textsf{Impre}_{\textsf{Var}}(n_2,{*}\textsf{ret})$ through inter-procedural edges so that TraceIntra can trace the imprecise return values from $n_1$ or $n_2$ in the callees as next targets to trace further. Note that we omitted the rules that generate precise tracing nodes $\textsf{Pre}_{\textsf{Var}}$ when the variables have precise values. In such cases, algorithm 1 stops tracing the imprecision.

## 5 Node/Edge Patterns in Tracing Graphs

Now, we present common node/edge patterns in tracing graphs to automatically detect main causes of frequent imprecision problems in analyzing JavaScript programs. Since our technique traces analysis imprecision from points with imprecise results back to points with precise ones, all the leaf nodes of tracing graphs have either precise results or imprecise results at the entry of the global scope. By inspecting such leaf nodes of a given tracing graph $G$, we identify four common patterns that represent main imprecision causes as follows:

$$\{\ \langle t,p\rangle \mid \langle N,E\rangle = G \ \land\ t \in N \ \land\ p = \textsc{Pattern}(t,E) \land\ p \neq \textsf{Unknown}\ \}$$

where $\text{Pattern}(t, E)$ defined in figure 8 detects imprecision patterns for a tracing node $t$ using a set of tracing edges $E$ in a tracing graph $G$.

**Function Pattern** In the first rule, the Function pattern denotes when precise values from callers are merged at the entry of a callee function. It simply checks if all successors of a tracing node that represents an imprecise value at a function entry are tracing nodes that represent precise values at function calls.

**Loop Pattern** In the second, the Loop pattern represents precision loss at a loop head. The predicate $t \rightsquigarrow t$ denotes that there exists a path from $t$ to $t$ in the tracing graph; if a tracing graph has a path from an imprecise node to itself and one of its successors has a precise value, Pattern concludes that the imprecision is due to the loop. Since we already presented the Function and Loop patterns with examples in section 2.2, we present the other two patterns—HeapClone and Model— in more detail in this section.

**HeapClone Pattern** Third, HeapClone is due to weak-update of object properties. For example:

```
1 function f(x) { return { foo : x }; }
2 var o1 = f(function f1() { });
3 var o2 = f(function f2() { });
4 o1.foo();
```

where function f creates a new object that has one property named foo, updates the property to the value of x passed by its caller, and returns the generated object to the callsite. Note that two callsites of f (lines 2 and 3) pass two functions f1 and f2 as the argument values of x, respectively. Then, f returns two different objects for o1 and o2 at each callsite. However, because static analyzers often abstract concrete objects on the same allocation site as one summary, the analysis values of o1 and o2 point to the same summary of two different concrete objects generated from different calling contexts with a weak-updated abstract value for o1.foo. Thus, analyzers conclude that both f1 and f2 may be called at o1.foo(). The following simplified tracing graph shows the pattern for the example code:

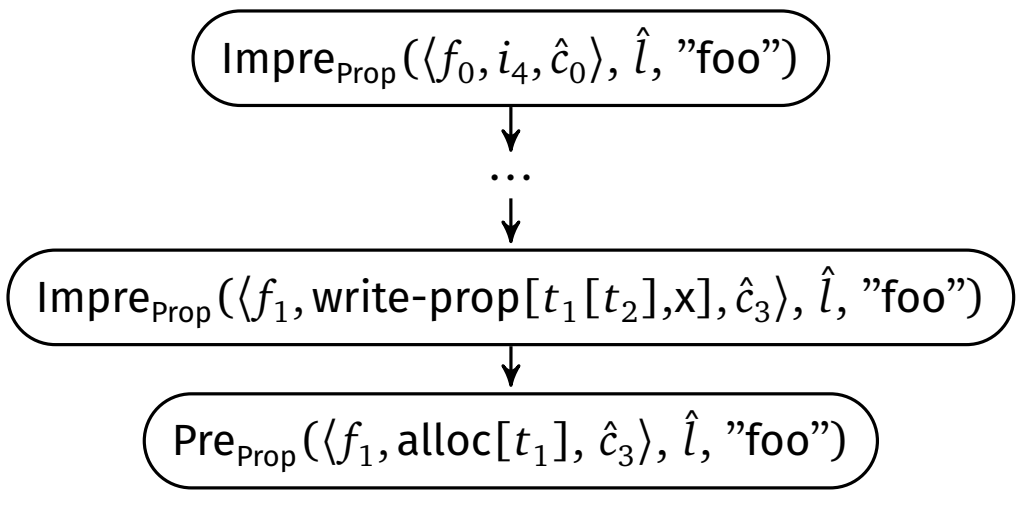

where $f_0$ and $f_1$ denote the function ids for the global scope and the function f, $\hat{c}_0$ and $\hat{c}_3$ denote the calling contexts of the global scope and the function call at line 3, $i_4$ denotes the call instruction at line 4, and $\hat{l}$ denotes the location of the generated object from f. Starting from the node representing that the abstract value of the foo property of the object at $\hat{l}$ is imprecise at line 4, our technique traces the imprecision back to the definition site of foo at line 1: write-prop[$t_1$[$t_2$],x] where the analysis values of $t_1$ and $t_2$ are $\hat{l}$ and “foo”. TraceIntra keeps tracing since $t_1$ and $t_2$ have precise values but $t_1[t_2]$ has an imprecise value. Finally, it reaches at the node where the foo property becomes imprecise after updating it. Pattern detects this pattern by identifying a tracing node that represents a property update with $\mathsf{Impre}_{\mathsf{Prop}}$, and its successor node with $\mathsf{Pre}_{\mathsf{Prop}}$.

While the example tracing graph shows the HeapClone pattern with write-prop, alloc that makes all properties absent also can generate this pattern using the same process. Consider the following example:

```
1 function f(b) {
2     var x = { };
3     if(b) x.foo = function f1() { };
4     return x;
5 }
6 var o1 = f(true);
7 var o2 = f(false);
8 o1.foo();
```

A function f generates an object (line 2), and returns the object having the property foo if an argument b is true or an empty object, otherwise; o1 points to an object that contains the property foo with the function object f1, and o2 points to an empty object. Then, o1.foo (line 8) calls the function f1. A simple static analysis of this program will report that o1 and o2 point to the same abstract object $\hat{l}$ generated at line 2, and the abstract object may or may not have the property foo. When our technique traces an imprecision for o1.foo (line 8), it generates the following simplified tracing graph:

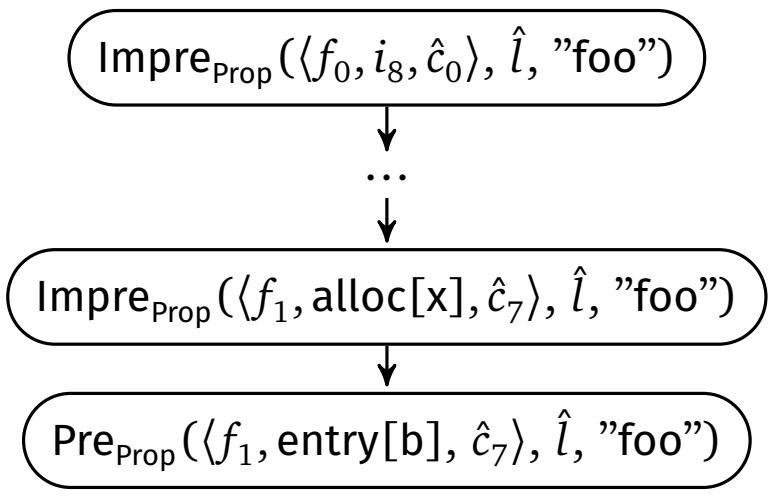

Since the property foo of $\hat{l}$ is imprecise at line 8, our technique traces it back to alloc[x] of the function f called at line 7. Because the abstract object $\hat{l}$ generated by alloc[x] abstracts two concrete objects, an empty object and the object with the property foo, the analysis performs a weak-update on $\hat{l}$ when it generates an object at line 2, which makes the abstract object may or may not have the property. Using our technique, we can find out that the property foo of $\hat{l}$ is precise before alloc[x] of the second f call and we can conclude that the weak-update is the main causes of imprecision. Pattern detects it as the HeapClone pattern.

**Model Pattern** The fourth pattern is Model that denotes imprecise values from imprecise models for built-in functions and objects. While static analyzers often support abstract models for built-in functions and objects like Math.random [25], imprecise models can be main causes of analysis imprecision. For example, consider the following partial tracing graph that shows the Model pattern for built-in functions:

$$\mathsf{Impre}_{\mathsf{Var}}(\langle f, \mathsf{builtin}[c, x_{\mathsf{this}}, x_{\mathsf{args}}], _\rangle, *\mathsf{ret})$$
$$\downarrow$$
$$\mathsf{Pre}_{\mathsf{Var}}(\langle f, \mathsf{entry}, _\rangle, x_{\mathsf{args}})$$

While the return value of the built-in function $f$ is imprecise, the abstract value for $x_{\mathsf{args}}$ used to compute the result of the built-in function is precise at the entry of $f$. Thus, we can conclude that the cause of the imprecision is the imprecise abstract model of $\mathsf{builtin}[c, x_{\mathsf{this}}, x_{\mathsf{args}}]$. In addition, the following shows the pattern for built-in objects:

$$\mathsf{Impre}_{\mathsf{Prop}}(\langle _, _, _\rangle, \hat{l}, s)$$
$$\downarrow$$
$$\mathsf{Impre}_{\mathsf{Prop}}(\langle f_0, \mathsf{entry}, \hat{c}_0\rangle, \hat{l}, s)$$

When the imprecision reaches the entry of the global scope, we can conclude that the imprecision is due to an imprecise initial abstract model for the property $s$ of the built-in object $\hat{l}$, which PATTERN can detect.

## 6 Experimental Evaluation

We implemented the technique on top of SAFE [17], an open-source, flow- and context-sensitive JavaScript analyzer, and evaluate it according to the following:

RQ1. **Effectiveness:** For given imprecise variables, does the technique effectively identify analysis imprecision?

RQ2. **Usefulness of patterns:** which patterns are useful in reducing the analysis imprecision?

RQ3. **Cost of graph building:** How long does it take to build tracing graphs?

RQ4. **Cost of imprecise cause tracing:** How many tracing graph edges do we need to backtrack to find the main causes of the imprecision?

### 6.1 Evaluation Methods

We evaluated the technique with 17 subjects: 12 targets that load jQuery versions 1.0∼1.11 and five from the test suite of JSAI, an open-source JavaScript analysis framework [13]. For precision measure, we consider a callsite whose analysis result is multiple function calls as an Imprecise Call (IC). Since the literature [1, 23, 24, 31, 33] showed that improving analysis precision in calls dramatically improves analysis scalability, we apply our technique to reduce the number of ICs.

For each target, we first analyze it context-insensitively (Base) with the timeout of 30 minutes, and identify reachable but imprecise callsites. Then, the variables that have an abstract value approximating multiple function values at those callsites become input variables for our technique to produce a tracing graph. The technique stops the analysis as soon as one of the input variables gets multiple function values;

it then produces a tracing graph to automatically detect the common node/edge patterns.

When it detects imprecision via node/edge patterns, we check if the patterns correctly indicate analysis imprecision according to the following (Selective):

- Function: Apply 10-CFA distinguishing different calling contexts with the maximum 10-length of a call history to the imprecise function calls.
- Loop: Apply a loop-sensitive analysis that distinguishes loop iterations to the loops with imprecision; we apply ⟨ 10, 100, 10 ⟩-LSA [24], which distinguishes up to 100 iterations for each of the maximum 10-nested loops with 10-CFA.
- HeapClone: Apply the heap cloning [16] technique that distinguishes different objects from different contexts to the imprecise object allocation sites.
- Model: Change the values of built-in object properties or built-in function results to random constant values. While it does not preserve the semantics, it simply checks whether this pattern actually indicates the analysis imprecision.

By selectively applying the techniques above only to certain program points, we can get required analysis precision at those points incurring only modest performance overhead while applying them to all program points could lead to huge overhead. We use the notion Trial and one trial represents the process from Static Analyzer to Pattern Identifier described in figure 1. In a trial, the technique analyzes the target program with the initial configuration and identifies an imprecise variable among the input variables. Then, it generates a tracing graph for the imprecise variable, and discovers a new configuration that can resolve analysis imprecision detected by the patterns above. We repeat such trials with the same input variables until the technique cannot provide any new information to detect imprecision problems within the timeout of 30 minutes.

We performed the experiments on a Linux x64 machine with 4.0 GHz Intel Core i7 CPU and 32 GB memory, and report an average of 30 runs for each experiment. We compare Selective with $\text{SAFE}_{\text{LSA}}$ [24] with the most precise configurations by applying loop-sensitivity to all loops and heap cloning to both heuristically selected allocation sites and identified ones by Selective.

### 6.2 Experimental Results

The experimental results for RQ1 are in table 1, RQ2 and RQ3 are in table 2, and RQ4 are in table 3.

#### 6.2.1 RQ1. Effectiveness

Table 1 shows whether the technique correctly identifies the analysis imprecision sources. The time and IC columns show analysis time in seconds and the numbers of ICs, respectively. When Base and $\text{SAFE}_{\text{LSA}}$ fail with the timeout, we show the numbers of all the reachable ICs during analysis in parentheses. For Selective, while Trial shows the numbers of trials to reduce the ICs, time and a.time show the analysis time for the final trial and accumulated analysis time for all trials, respectively, and non-parenthesized ICs and parenthesized ICs show the ICs for the final trial and all trials, respectively.

**Table 1** Analysis cost and precision

| Target | | Base | | $SAFE_{LSA}$ | | Selective | | | |
|---|---|---|---|---|---|---|---|---|---|
| Name | Loc | time (s) | IC | time (s) | IC | time (s) | IC | Trial | a.time (s) |
| jQuery 1.0 | 1,194 | ✗ | (114) | 1.8 | 0 | 1.9 | 0( 2) | 4 | 4.5 |
| jQuery 1.1 | 1,437 | ✗ | (77) | 2.4 | 0 | 1.3 | 0( 2) | 4 | 3.0 |
| jQuery 1.2 | 1,942 | ✗ | (503) | 10.0 | 0 | 7.6 | 0( 2) | 3 | 8.9 |
| jQuery 1.3 | 2,849 | ✗ | (34) | ✗ | (19) | 38.1 | 0( 6) | 7 | 183.9 |
| jQuery 1.4 | 4,143 | ✗ | (54) | 32.9 | 4 | 48.5 | 0( 4) | 6 | 100.5 |
| jQuery 1.5 | 5,347 | ✗ | (34) | ✗ | (778) | 67.6 | 0( 5) | 6 | 213.3 |
| jQuery 1.6 | 5,852 | ✗ | (775) | ✗ | (80) | 67.6 | 0( 4) | 6 | 236.0 |
| jQuery 1.7 | 6,409 | ✗ | (773) | 111.7 | 7 | 62.0 | 1( 7) | 13 | 566.7 |
| jQuery 1.8 | 6,355 | ✗ | (52) | 307.7 | 15 | 26.2 | 1(11) | 17 | 182.7 |
| jQuery 1.9 | 6,550 | ✗ | (47) | 203.0 | 15 | 25.9 | 1(12) | 18 | 197.9 |
| jQuery 1.10 | 6,632 | ✗ | (100) | 98.4 | 16 | 27.2 | 1(12) | 18 | 211.7 |
| jQuery 1.11 | 6,866 | ✗ | (45) | 62.4 | 12 | 28.8 | 1(12) | 18 | 233.1 |
| action | 2,294 | 50.1 | 24 | 3.7 | 0 | 12.9 | 0(11) | 15 | 47.7 |
| aggregate | 2,331 | 745.9 | 36 | 239.3 | 18 | 30.9 | 0(14) | 27 | 239.8 |
| dictionary | 2,420 | 610.3 | 26 | 4.0 | 0 | 47.0 | 0( 9) | 20 | 414.1 |
| enumerable | 2,374 | 1,034.9 | 43 | 13.7 | 0 | 42.0 | 0(12) | 20 | 179.1 |
| functional | 2,301 | 867.8 | 53 | 5.8 | 0 | 5.9 | 1(25) | 41 | 1782.5 |

The Base analyses could not finish analyzing any jQuery versions within the timeout, while Selective finished all the targets. Selective resolved all ICs in 11 targets and left one IC unresolved in six targets at the final trials. For instance, to analyze jQuery version 1.11, the technique encountered 12 ICs during 18 trials within 233.1 seconds, and finally it finished the analysis in 28.8 seconds with one IC. For functional in the benchmarks, Base generates 53 ICs while Selective generates one IC; it encountered 25 ICs during 41 trials within 1782.5 seconds and finally finished the analysis in 5.9 seconds.

The last trial of Selective that analyzes programs with accumulated analysis configurations by our technique finishes 11 targets out of 17 significantly faster showing higher precision than $SAFE_{LSA}$. For instance, while $SAFE_{LSA}$ finishes the analysis of jQuery 1.8 in 307.7 seconds with 15 ICs, Selective finishes it in 26.2 seconds with only one IC unresolved. Even for three targets that $SAFE_{LSA}$ could not finish analyzing within the timeout, Selective finished them within 2 minutes. While Selective significantly improves precision and scalability at the last trial, static analysis with iterative refinement of analysis configurations may be expensive. For example, functional spends 1782.5 seconds to discover main causes of 25 ICs during 41 trials while the last trial only takes 5.9 seconds. Nevertheless, our technique analyzes six out of 17 targets faster than $SAFE_{LSA}$ with more precise results.

*Answer to RQ1.* The technique helped correctly identify and resolve 144 out of total 150 [a] (96 %) imprecision problems in 17 JavaScript applications.

[a] (B-A)/B where A and B are the sums of the non-parenthesized and parenthesized numbers in the IC column under Selective of table 1, respectively.

**Table 2** Experimental results for RQ2 and RQ3

| Name | RQ2 | | | | | RQ3 | | | |
|---|---|---|---|---|---|---|---|---|---|
| | Pattern | | | | | Node Size | | | Time (s) |
| | F | L | H | M | Total | Min | Max | Avg | Avg |
| jQuery 1.0 | 2 | 1 | 0 | 1 | 4 | 8 | 73 | 45 | 0.2 |
| jQuery 1.1 | 2 | 2 | 0 | 1 | 5 | 8 | 101 | 52 | 0.2 |
| jQuery 1.2 | 2 | 2 | 0 | 1 | 5 | 8 | 106 | 57 | 0.2 |
| jQuery 1.3 | 5 | 3 | 0 | 1 | 9 | 8 | 150 | 78 | 0.2 |
| jQuery 1.4 | 4 | 2 | 0 | 1 | 7 | 8 | 162 | 57 | 0.3 |
| jQuery 1.5 | 5 | 5 | 0 | 1 | 11 | 17 | 344 | 136 | 0.3 |
| jQuery 1.6 | 4 | 5 | 0 | 2 | 11 | 8 | 325 | 145 | 0.6 |
| jQuery 1.7 | 7 | 8 | 0 | 2 | 17 | 6 | 450 | 141 | 0.5 |
| jQuery 1.8 | 8 | 6 | 4 | 3 | 21 | 6 | 424 | 113 | 0.3 |
| jQuery 1.9 | 8 | 6 | 5 | 2 | 21 | 6 | 588 | 128 | 0.3 |
| jQuery 1.10 | 9 | 9 | 5 | 2 | 25 | 6 | 713 | 136 | 0.3 |
| jQuery 1.11 | 8 | 9 | 5 | 1 | 23 | 6 | 727 | 131 | 0.4 |
| action | 9 | 2 | 4 | 0 | 15 | 6 | 483 | 100 | 0.1 |
| aggregate | 15 | 1 | 11 | 0 | 27 | 6 | 480 | 63 | 0.1 |
| dictionary | 15 | 5 | 7 | 0 | 27 | 6 | 406 | 66 | 0.1 |
| enumerable | 16 | 1 | 3 | 0 | 20 | 6 | 261 | 69 | 0.1 |
| functional | 12 | 1 | 17 | 0 | 30 | 6 | 16,522 | 3,758 | 0.7 |

### 6.2.2 RQ2. Usefulness of Patterns

We evaluated which node/edge patterns are effective in reducing analysis imprecision with the number of each pattern appearing in tracing graphs. RQ2 in table 1, F, L, H, and M stand for Function, Loop, HeapClone, and Model patterns, respectively. For instance, in the case of jQuery version 1.10, during 18 trials, our technique detected 25 patterns, among which nine Function and Loop, five HeapClone, and two Model patterns appeared.

The most common pattern is precision losses at function entries (47 %). For jQuery libraries, loops are the second most pattern (24 %), and HeapClone is the second for the other benchmarks (22 %). We observed that all analyses of jQuery libraries have the Model pattern. Our manual inspection revealed that statically indeterminate return values from built-in functions like new Date() and Math.random() and browser-specific data like navigator.userAgent and document.readyState were the main causes of the imprecision.

*Answer to RQ2*. The Function pattern is the most effective pattern to automatically find the causes of analysis imprecision in both jQuery libraries and benchmarks.

### 6.2.3 RQ3. Cost of Graph Building

We evaluated the cost of building tracing graphs by measuring the minimum, maximum, and average numbers of nodes in tracing graphs, and the average time to build tracing graphs as shown in table 2. For each target, we considered all tracing graphs that the technique built during all the trials. The minimum and average numbers of nodes are six and 310.6, respectively, and the average time to build tracing graphs is less than 1 second. The outlier functional has the maximum 16,522 nodes, and it took the longest time to build tracing graphs. We found that the technique traced about 26 nodes for each calling context on average from 633 different calling contexts, which leads to many tracing nodes incurring large performance overhead.

**Table 3** Cost of tracing

| Length | # of shortest paths | | | | |
|---|---|---|---|---|---|
| | F | L | H | M | Total |
| 0∼ 9 | 56 | 17 | 0 | 0 | 73 |
| 10∼19 | 38 | 26 | 12 | 3 | 79 |
| 20∼29 | 17 | 7 | 18 | 3 | 45 |
| 30∼39 | 11 | 7 | 3 | 0 | 21 |
| 40∼49 | 7 | 0 | 3 | 2 | 12 |
| 50∼59 | 0 | 5 | 4 | 1 | 10 |
| 60∼69 | 1 | 4 | 3 | 0 | 8 |
| 70∼79 | 0 | 3 | 1 | 0 | 8 |
| 80∼89 | 1 | 0 | 2 | 0 | 3 |
| 90∼99 | 0 | 0 | 1 | 0 | 1 |
| ≥100 | 0 | 1 | 12 | 9 | 22 |
| Total | 131 | 70 | 59 | 18 | 278 |
| Average Length | 16.1 | 25.2 | 63.9 | 104.2 | 87.2 |

*Answer to RQ3.* Building tracing graphs incurs only negligible performance overheads; it took less than 1 second to build a tracing graph on average.

#### 6.2.4 RQ4. Cost of Imprecise Cause Tracing

To understand the cost of debugging analysis imprecision, we measured the numbers of tracing graph edges from imprecise input variables to their corresponding imprecision patterns. Out of 227 tracing graphs that the technique constructed throughout the evaluation, we collected 278 shortest paths. We call the length of such a path a *tracing graph length*, and table 3 shows the tracing graph lengths of 278 paths. It shows the approximate amount of work required to manually investigate tracing graphs to figure out main causes of analysis imprecision. The average tracing graph length for all patterns is 87.2; while the average tracing graph lengths for `Function` and `Loop` are 16.1 and 25.2, respectively, those of `HeapClone` and `Model` are 63.9 and 104.2. Thus, debugging `HeapClone` and `Model` patterns may take more time than `Function` and `Loop`. We also found that the shortest path lengths of about 92 % of tracing graphs are less than 100.

*Answer to RQ4.* The number of tracing graph edges to backtrack to identify an imprecision cause is modest; the average number is 87.2.

### 6.3 Threats to Validity

One possible threat to the validity of our evaluation is the experimentation setting for inputs to our technique. Even though the technique can trace imprecision causes of any variables with any imprecise values when they are given as inputs, our experiments considered only the variables that may have multiple function values at callsites as inputs. When other variables with different imprecise values are given as inputs, the evaluation results may show different patterns.

Another validity threat is the representativeness of our evaluation targets. While the evaluation targets are also used in various previous studies [1, 23, 24, 31, 33], they do not cover the wide spectrum of JavaScript applications. Statically analyzing them may impose other challenges, and our future work includes applying the technique to more various program domains.

## 7 Related Work

**Manual investigation of analysis imprecision causes** Researchers [1, 24, 31] have diagnosed main causes of scalability problems in statically analyzing JavaScript libraries as precision losses in loops by manual investigation. They presented various loop specialization techniques to improve precision in analyzing loops. Our technique can accelerate their research processes by replacing the manual investigation phase with automatic detection.

**Forward propagation with predefined root causes** Several techniques [9, 33] use forward-propagation analysis from predefined sets of all possible root causes of analysis imprecision to localize main root causes among them. A process of root cause localization and remediation in [33] regards all variables and property accesses as possible root causes for overall analysis imprecision. Guyer and Lin's work [9] keeps track of data flows from all the variables whose values are merged by flow-insensitivity or context-insensitivity, and maintains information from all predefined root causes. On the contrary, our technique *backtracks* the main sources that lead to imprecise values for specific variables keeping only relevant information for the imprecision. Note that analysis imprecision may be due to various reasons like implementation bugs or imprecise transfer functions besides predefined root causes. Our technique can trace such causes as well.

**CEGAR approach** The counterexample guided abstraction refinement (CEGAR) approach [5] based on model checking was devised to reduce the cost of exploring large search space exhaustively. From a cheap initial abstraction of the input space, it iteratively finds a spurious counterexample that disproves a given property but can never happen in any concrete execution. Then, it refines the current abstraction so that the example cannot happen in subsequent abstractions. Several work applied this approach to static analysis of Java programs specified in Datalog [19, 35]. Their techniques identify parts of programs that can reach to unproven queries and efficiently find a cheap abstraction that can remedy the unproven queries among a set of possible abstractions. Our technique, on the other hand, focuses on finding direct imprecision causes, depending on which, it is possible to apply not only more suitable abstraction but also other methods such as loop unrolling and user inputs to improve analysis precision.

**Selective analysis** Recent studies on selective analysis [21, 28, 32] proposed techniques to selectively apply context-sensitivity only to some function calls that need higher precision. One approach uses cheap pre-analyses to extract heuristically chosen characteristics of each function such as the number of callsites and sizes of points-to sets for pointer variables [28, 32]. Then, subsequent analyses apply different context-sensitivities to functions depending on their characteristics. Another approach applies selective analysis to improve analysis scalability in proving queries related to buffer overflow [21]. At a pre-analysis stage, it uses a simple abstract value domain only with two elements that denote positive and unknown signs of buffer indices. With full context-sensitivity, it identifies call contexts where given queries are likely to be proven with the positive sign of buffer indices. Then, in a subsequent main analysis with a more complex domain that has an enough expressive power to prove the target queries, it selectively applies context-sensitivities only to the identified call contexts to improve the analysis scalability. We also used a selective analysis to programs to check whether our solutions, which were selectively applied to the node/edge patterns that our technique identified, really resolve the imprecision problems. We believe that our technique can be combined with various selective analyses to determine which analysis techniques to use and where to apply them.

**Backward analysis** Several works on backward analysis [2, 3, 4, 7] check the validity of given queries. Their techniques keep applying transfer functions that infer pre-conditions backward from given post-conditions until they find any contradictory condition that refutes the queries. Demand-driven analysis [8, 10, 22, 29, 30, 34] also performs a backward analysis, by which it finds relevant parts of programs to prove given queries and applies context-sensitivity to the functions that require higher precision in the program parts. Their goal of using backward analysis is different from ours in that they use a backward analysis to find contradictory pre-conditions or relevant program parts for proving some queries, while we use it to find major causes of analysis imprecision from some preceding forward analysis results.

**Value flow patterns** Zipper [18] uses value flow patterns that incur precision losses in context-insensitive analysis to identify methods that require higher analysis precision. Their technique discovers patterns for object flows from parameters to return sites through assignments, field accesses, and method calls that transfer analysis imprecision. Similarly, Guyer and Lin [9] presented object flow patterns via assignments that also may incur precision losses in context-insensitive and flow-insensitive analysis. More specifically, their technique identifies two assignment patterns: parameter assignments that merge points-to sets from different call sites and a set of assignments that updates the same memory address with different points-to sets. When the techniques identify such value flow patterns, they apply context-sensitivity or flow-sensitivity to the methods that contain the patterns for getting higher precision.

Our technique traces flows of not only objects but also primitive values to identify main causes of imprecision. In JavaScript static analysis, imprecision of indices used in loops and property names can lead to immense precision losses and scalability problems [1, 15, 24, 31]. Because flows of points-to information are not enough to identify such imprecision, our technique traces value flows of programs, and provides patterns to discover main causes of imprecision.

## 8 Conclusion

We presented a novel technique to trace analysis imprecision in JavaScript static analysis using tracing graphs that illustrate trace information from imprecise program points to the staring points of the imprecision. We also presented four common node/edge patterns in tracing graphs to enable automatic detection of the four common main causes of analysis imprecision such as precision losses in loops. We experimentally demonstrated that our technique can find and help resolving 96 % (144 out of 150) of the main causes of imprecision problems in 17 JavaScript applications by only automatic detection.

**Acknowledgements** This work has received funding from National Research Foundation of Korea (NRF) (Grants NRF-2017R1A2B3012020 and 2017M3C4A7068177).

## References

[1] Esben Andreasen and Anders Møller. "Determinacy in static analysis for jQuery". In: *Proceedings of the ACM International Conference on Object Oriented Programming Systems Languages and Applications*. 2014. DOI: 10.1145/2660193.2660214.

[2] Sam Blackshear, Bor-Yuh Evan Chang, and Manu Sridharan. "Selective control-flow abstraction via jumping". In: *Proceedings of the ACM International Conference on Object Oriented Programming Systems Languages and Applications*. 2015. DOI: 10.1145/2814270.2814293.

[3] Sam Blackshear, Bor-Yuh Evan Chang, and Manu Sridharan. "Thresher: Precise refutations for heap reachability". In: *Proceedings of the ACM SIGPLAN Conference on Programming Language Design and Implementation*. 2013. DOI: 10.1145/2499370.2462186.

[4] Satish Chandra, Stephen J. Fink, and Manu Sridharan. "Snugglebug: A powerful approach to weakest preconditions". In: *Proceedings of the ACM SIGPLAN Conference on Programming Language Design and Implementation*. 2009. DOI: 10.1145/1542476.1542517.

[5] Edmund Clarke, Orna Grumberg, Somesh Jha, Yuan Lu, and Helmut Veith. "Counterexample-guided abstraction refinement". In: *Proceedings of the International Conference on Computer Aided Verification*. 2000. DOI: 10.1007/10722167_15.

[6] Patrick Cousot and Radhia Cousot. "Abstract Interpretation: A unified lattice model for static analysis of programs by construction or approximation of fixpoints". In: *Proceedings of the ACM SIGACT-SIGPLAN Symposium on Principles of Programming Languages*. 1977. DOI: 10.1145/512950.512973.

[7] Patrick Cousot, Radhia Cousot, and Francesco Logozzo. "Precondition inference from intermittent assertions and application to contracts on collections". In: *Proceedings of the International Conference on Verification, Model Checking, and Abstract Interpretation*. 2011. DOI: 10.1007/978-3-642-18275-4_12.

[8] Leandro Facchinetti, Zachary Palmer, and Scott F. Smith. "Relative store fragments for singleton abstraction". In: *Proceedings of the International Conference on Static Analysis*. 2017. DOI: 10.1007/978-3-319-66706-5_6.

[9] Samuel Z. Guyer and Calvin Lin. "Client-driven pointer analysis". In: *Proceedings of the International Conference on Static Analysis*. 2003. DOI: 10.1007/3-540-44898-5_12.

[10] Susan Horwitz, Thomas Reps, and Mooly Sagiv. "Demand interprocedural dataflow analysis". In: *Proceedings of the ACM SIGSOFT Symposium on Foundations of Software Engineering*. 1995. DOI: 10.1145/222124.222146.

[11] IBM Research. *WALA*. URL: http://wala.sf.net (visited on 2019-01-28).

[12] Simon Holm Jensen, Anders Møller, and Peter Thiemann. "Type analysis for JavaScript". In: *Proceedings of the International Conference on Static Analysis*. 2009. DOI: 10.1007/978-3-642-03237-0_17.

[13] Vineeth Kashyap, Kyle Dewey, Ethan A. Kuefner, John Wagner, Kevin Gibbons, John Sarracino, Ben Wiedermann, and Ben Hardekopf. "JSAI: A static analysis platform for JavaScript". In: *Proceedings of the ACM SIGSOFT International Symposium on Foundations of Software Engineering*. 2014. DOI: 10.1145/2635868.2635904.

[14] Gary A. Kildall. "A unified approach to global program optimization". In: *Proceedings of the ACM SIGACT-SIGPLAN Symposium on Principles of Programming Languages*. 1973. DOI: 10.1145/512927.512945.

[15] Yoonseok Ko, Xavier Rival, and Sukyoung Ryu. "Weakly sensitive analysis for unbounded iteration over JavaScript objects". In: *Proceedings of the Asian Symposium on Programming Languages and Systems*. 2017. DOI: 10.1007/978-3-319-71237-6_8.

[16] Chris Lattner, Andrew Lenharth, and Vikram Adve. "Making context-sensitive points-to analysis with heap cloning practical for the real world". In: *Proceedings of the ACM SIGPLAN Conference on Programming Language Design and Implementation*. 2007. DOI: 10.1145/1250734.1250766.

[17] Hongki Lee, Sooncheol Won, Joonho Jin, Junhee Cho, and Sukyoung Ryu. "SAFE: Formal specification and implementation of a scalable analysis framework for ECMAScript". In: *Proceedings of the International Workshop on Foundations of Object-Oriented Languages*. 2012.

[18] Yue Li, Tian Tan, Anders Møller, and Yannis Smaragdakis. "Precision-guided context sensitivity for pointer analysis". In: *Proceedings of the ACM International Conference on Object Oriented Programming Systems Languages and Applications*. 2018. DOI: 10.1145/3276511.

[19] Percy Liang and Mayur Naik. "Scaling abstraction refinement via pruning". In: *Proceedings of the ACM SIGPLAN Conference on Programming Language Design and Implementation*. 2011. DOI: 10.1145/1993498.1993567.

[20] Magnus Madsen and Esben Andreasen. "String analysis for dynamic field access". In: *Proceedings of the International Conference on Compiler Construction*. 2014. DOI: 10.1007/978-3-642-54807-9_12.

[21] Hakjoo Oh, Wonchan Lee, Kihong Heo, Hongseok Yang, and Kwangkeun Yi. "Selective context-sensitivity guided by impact pre-analysis". In: *Proceedings of the ACM SIGPLAN Conference on Programming Language Design and Implementation*. 2014. DOI: 10.1145/2594291.2594318.

[22] Zachary Palmer and Scott F. Smith. "Higher-order demand-driven program analysis". In: *Proceedings of the European Conference on Object-Oriented Programming*. 2016. DOI: 10.4230/LIPIcs.ECOOP.2016.19.

[23] Changhee Park, Hyeonseung Im, and Sukyoung Ryu. "Precise and scalable static analysis of jQuery using a regular expression domain". In: *Proceedings of the Symposium on Dynamic Languages*. 2016. DOI: 10.1145/2989225.2989228.

[24] Changhee Park, Hongki Lee, and Sukyoung Ryu. "Static analysis of JavaScript libraries in a scalable and precise way using loop sensitivity". In: *Software: Practice and Experienc*. 2018. DOI: 10.1002/spe.2552.

[25] Changhee Park, Sooncheol Won, Joonho Jin, and Sukyoung Ryu. "Static analysis of JavaScript web applications in the wild via practical DOM modeling". In: *Proceedings of the 30th IEEE/ACM International Conference on Automated Software Engineering*. 2015. DOI: 10.1109/ASE.2015.27.

[26] Micha Sharir and Amir Pnueli. "Two approaches to interprocedural data Flow analysis". In: *Program Flow Analysis: Theory and Applications, Chapter 7*. Prentice-Hall, 1981. ISBN: 978-0137296811.

[27] Olin Shivers. "Control flow analysis in Scheme". In: *Proceedings of the ACM SIGPLAN Conference on Programming Language Design and Implementation*. 1988. DOI: 10.1145/53990.54007.

[28] Yannis Smaragdakis, George Kastrinis, and George Balatsouras. "Introspective analysis: Context-sensitivity, across the board". In: *Proceedings of the ACM SIGPLAN Conference on Programming Language Design and Implementation*. 2014. DOI: 10.1145/2594291.2594320.

[29] Johannes Späth, Lisa Nguyen Quang Do, Karim Ali, and Eric Bodden. "Boomerang: Demand-driven flow- and context-sensitive pointer analysis for Java". In: *Proceedings of the European Conference on Object-Oriented Programming*. 2016. DOI: 10.4230/LIPIcs.ECOOP.2016.22.

[30] Manu Sridharan and Rastislav Bodík. "Refinement-based context-sensitive points-to analysis for Java". In: *Proceedings of the ACM SIGPLAN Conference on Programming Language Design and Implementation*. 2006. DOI: 10.1145/1133981.1134027.

[31] Manu Sridharan, Julian Dolby, Satish Chandra, Max Schäfer, and Frank Tip. "Correlation tracking for points-to analysis of JavaScript". In: *Proceedings of the European Conference on Object-Oriented Programming*. 2012. DOI: 10.1007/978-3-642-31057-7_20.

[32] Shiyi Wei and Barbara G. Ryder. "Adaptive context-sensitive analysis for JavaScript". In: *Proceedings of the European Conference on Object-Oriented Programming*. 2015. DOI: 10.4230/LIPIcs.ECOOP.2015.712.

[33] Shiyi Wei, Omer Tripp, Barbara G. Ryder, and Julian Dolby. "Revamping JavaScript static analysis via localization and remediation of root causes of imprecision". In: *Proceedings of the ACM SIGSOFT International Symposium on Foundations of Software Engineering*. 2016. DOI: 10.1145/2950290.2950338.

[34] Dacong Yan, Guoqing Xu, and Atanas Rountev. "Demand-driven context-sensitive alias analysis for Java". In: *Proceedings of the International Symposium on Software Testing and Analysis*. 2011. DOI: 10.1145/2001420.2001440.

[35] Xin Zhang, Ravi Mangal, Radu Grigore, Mayur Naik, and Hongseok Yang. "On abstraction refinement for program analyses in Datalog". In: *Proceedings of the ACM SIGPLAN Conference on Programming Language Design and Implementation*. 2014. DOI: 10.1145/2594291.2594327.

## About the authors

**Hongki Lee** is a Ph.D. student at KAIST. Contact him at petitkan@kaist.ac.kr.

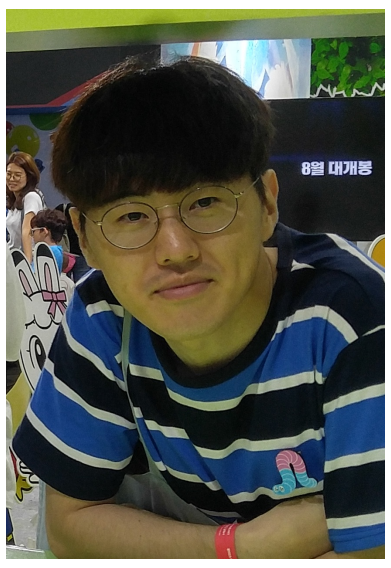

**Changhee Park** Contact him at changhee.park@kaist.ac.kr.

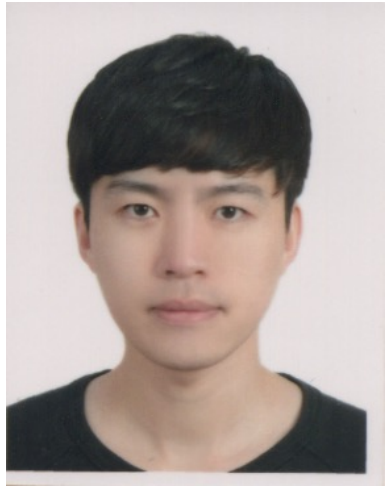

**Sukyoung Ryu** is an associate professor in the School of Computing at Korea Advanced Institute of Science and Technology (KAIST). Her research interests include programming languages, program analysis, and programming environment. Ryu received a Ph.D. in computer science from KAIST. She is a Member of the IEEE and ACM. Contact her at sryu.cs@kaist.ac.kr.

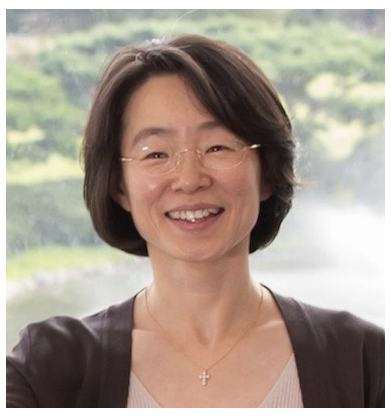